\documentclass{IEEEoj}
\usepackage{cite}
\usepackage{amsmath,amssymb,amsfonts}
\usepackage{algorithmic}
\usepackage{graphicx,color}
\usepackage{xcolor}
\usepackage{textcomp}
\def\BibTeX{{\rm B\kern-.05em{\sc i\kern-.025em b}\kern-.08em
    T\kern-.1667em\lower.7ex\hbox{E}\kern-.125emX}}
\def\OJlogo{\vspace{-10pt}\includegraphics[height=20pt]{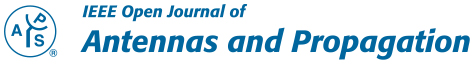}}
\begin{document}
\receiveddate{XX Month, XXXX}
\reviseddate{XX Month, XXXX}
\accepteddate{XX Month, XXXX}
\publisheddate{XX Month, XXXX}
\currentdate{XX Month, XXXX}
\doiinfo{OJAP.2026.XXXXXXX}

\title{\textsc{From Multi-Port Models to Cascade Structures: Optimization of Active Unilateral Stacked Intelligent Metasurfaces}}

\author{ANDREA ABRARDO\authorrefmark{1}, SENIOR MEMBER, IEEE, GIULIO BARTOLI\authorrefmark{1}, MEMBER, IEEE, MARCO DI RENZO\authorrefmark{2,3},FELLOW, IEEE, AND ALBERTO TOCCAFONDI\authorrefmark{1}, SENIOR MEMBER, IEEE}
\affil{Department of Information Engineering and Mathematics, University of Siena, Siena, Italy}
\affil{CNRS and CentraleSup\'elec, Institute of Electronics and Digital Technologies (IETR), Rennes, France}
\affil{King's College London, Centre for Telecommunications Research, London, UK}
\corresp{CORRESPONDING AUTHOR: Alberto Toccafondi (e-mail: alberto.toccafondi@unisi.it).}
\markboth{Preparation of Papers for \textsc{IEEE Open Journal of Antennas and Propagation}}{Author \textit{et al.}}

\begin{abstract}
This paper develops a multi-port S-parameter framework for the analysis and optimization of stacked intelligent metasurfaces (SIMs) with unilateral active interconnections.
By modeling each unit cell as a non-reciprocal two-port network, the resulting SIM exhibits a feed-forward structure that enables a recursive, cascade-like representation of the end-to-end transfer function while preserving electromagnetic accuracy.
Based on this model, we derive an efficient gradient-based optimization algorithm with reduced computational complexity compared to conventional reciprocal SIM architectures.
Numerical results, obtained from full-wave simulations, illustrate the trade-offs among inter-layer spacing, active gain, and SIM size in terms of channel diagonalization and achievable spectral efficiency.
\end{abstract}


\IEEEspecialpapernotice{(Invited Paper)}

\maketitle

\section{INTRODUCTION}
\IEEEPARstart{T}{he} continuous evolution of modern wireless communication systems demands innovative solutions to improve signal processing capabilities while maintaining energy efficiency. In this context, the concept of holographic multiple-input multiple-output (H-MIMO) communications \cite{Pizzo2020,Pizzo2022} has recently emerged as a promising paradigm for next-generation wireless systems. Unlike conventional beamforming architectures, H-MIMO relies on wave-domain processing, where the electromagnetic field is manipulated over a nearly continuous aperture to achieve highly efficient spatial multiplexing and wavefront control.
A practical realization of this concept is represented by reconfigurable holographic surfaces (RHSs) \cite{Deng2021RHS}, which implement holographic beamforming through densely packed tunable meta-elements distributed over a single planar layer. 

Building upon the same wave-domain processing principles, stacked intelligent metasurfaces (SIMs) have recently attracted significant attention as an advanced architecture capable of controlling electromagnetic wave propagation through the cascade of multiple transmitting reconfigurable intelligent surfaces (T-RISs) \cite{MDR_state_of_the_art}. Within this framework, SIMs can be regarded as a multi-layer realization of the broader H-MIMO paradigm, where wave-domain signal processing is performed through cascaded programmable metasurface layers. By integrating tunable meta-atoms across multiple layers, SIMs provide additional electromagnetic degrees of freedom for wave manipulation, enabling advanced wavefront shaping and programmable control of amplitude and phase while overcoming some of the spatial degrees-of-freedom limitations inherent to single-layer implementations \cite{An2023}. More recently, this concept has been further generalized to spherical stacked intelligent metasurfaces, enabling full-space wave-domain processing and opening new opportunities for future 6G communication systems \cite{Zhang2026SSIM}.

Recent studies have demonstrated the effectiveness of SIMs in a variety of applications, including electromagnetic beamforming \cite{MDR_state_of_the_art}, multi-stream MIMO communications with a reduced number of radio-frequency (RF) chains \cite{Hassan2024,An2023}, near-field communications \cite{Papazafeiropoulos2024a}, and sensing tasks such as direction-of-arrival estimation \cite{An2024b,An2024a}. Moreover, SIMs have been investigated in the context of integrated sensing and communications (ISAC), where they can jointly enhance communication and sensing performance \cite{Niu2024,SIM_ISAC,AntiJammingSIM,ISAC_MDR}.

Despite these promising capabilities, the design and optimization of SIMs remain challenging due to the lack of accurate electromagnetic models that can capture the complex interactions among layers and between the SIM and the surrounding environment. 

Most existing works rely on simplified models in which the SIM is described as a cascade of ideal transmitting arrays, where each layer applies a phase shift and the propagation between layers is modeled through simplified channel coefficients. While these models are appealing for system-level analysis, they neglect important electromagnetic interactions and do not accurately represent realistic SIM implementations.

On the other hand, more accurate modeling approaches based on multi-port network representations have been recently proposed in \cite{Nerini_Clerckx_SIM,AbrardoSIM2025}. In particular, the impedance-based framework introduced in \cite{AbrardoSIM2025} provides a rigorous electromagnetic description of SIMs, accounting for mutual coupling effects and enabling the development of optimization algorithms grounded in physical modeling.

In this paper, we bridge the gap between these two modeling approaches by considering a realistic multi-port network formulation, in which the interconnections between SIM layers are implemented through non-reciprocal (unilateral) active two-port networks.

This modeling choice is not only of theoretical interest,
but also practically motivated.
In \cite{Taravati2017}, a nonreciprocal metasurface is designed and
fabricated using a Surface-Circuit-Surface (SCS) architecture, in which
a unilateral transistor amplifier is placed between two antenna layers.
This structure realizes, at the unit-cell level, the same one-way
transmission mechanism modeled in this paper through the unilateral
two-port network. A programmable version of this concept, in which the
direction of nonreciprocal transmission can be controlled digitally via
integrated power amplifiers, is demonstrated in \cite{Ma2019}.
The extension to a multi-layer architecture is realized in \cite{Liu2022},
where a five-layer active metasurface array is used for wave-domain
processing. In the practical implementation of \cite{Liu2022}, the
combination of the unidirectionality of the amplifiers and the use of
microwave absorbing materials around the structure is shown to effectively
suppress electromagnetic coupling between non-adjacent layers.
This provides direct experimental evidence that the inter-layer isolation
assumption adopted in
Section~\ref{sec:SIM_model} of this paper can be closely approached in practice.

A preliminary version of this work was presented in \cite{AbrardoIceaa2025}, 
where a multi-port network model for SIMs with unilateral interconnections 
was developed by using a $Z$-parameters network representation, together with a corresponding 
optimization framework.

Compared to \cite{AbrardoIceaa2025}, the present paper introduces several 
substantial extensions. In particular, we develop a novel S-parameter-based 
formulation, which enables a direct connection with electromagnetic 
full-wave simulations, and we show that the unilateral architecture induces 
a feed-forward propagation mechanism that leads to a recursive 
cascade representation of the SIM transfer function.

This modeling choice is of particular interest because it allows us to retain 
the physical accuracy of the multi-port electromagnetic description while, 
at the same time, inducing a feed-forward structure in the signal propagation 
across the SIM layers.

As a result, the proposed model naturally leads to a cascade-like representation of the SIM, which resembles the simplified models commonly adopted in the communication-theoretic literature, despite being derived from a fully electromagnetic multi-port formulation. This shows that the classical cascade models can be interpreted as a particular case of a more general and physically grounded framework when unilateral devices are employed.

Building on this model, we develop an optimization algorithm specifically tailored to unilateral SIM architectures. Thanks to the feed-forward structure induced by the unilateral two-port networks, the proposed algorithm significantly reduces the computational complexity compared to both general electromagnetic collaborative objects (ECOs) and conventional SIMs with reciprocal passive loads.

We also provide new numerical results that offer additional insights into 
the trade-offs among inter-layer spacing, active gain, and SIM size.

The main contributions of this paper can be summarized as follows:
\begin{itemize}
    \item We develop a multi-port S-parameter model for SIMs that explicitly accounts for unilateral active interconnections between layers.
    \item We show that the unilateral structure induces a feed-forward propagation mechanism, which enables a recursive representation of the SIM transfer function.
    \item We derive an efficient optimization algorithm that exploits this structure, achieving a substantial reduction in computational complexity.
    \item We provide numerical results that highlight the trade-offs among inter-layer spacing, active gain, and SIM size, and their impact on channel diagonalization and achievable capacity.
\end{itemize}

Overall, the proposed framework provides a unified perspective that combines physical accuracy and algorithmic efficiency, offering a practical tool for the design and optimization of SIM-based communication systems.

\section{Multi-port S-Parameters SIM Model}
\label{sec:SIM_model}

\begin{figure}[h!]
	\centering
	\includegraphics[width=\columnwidth]{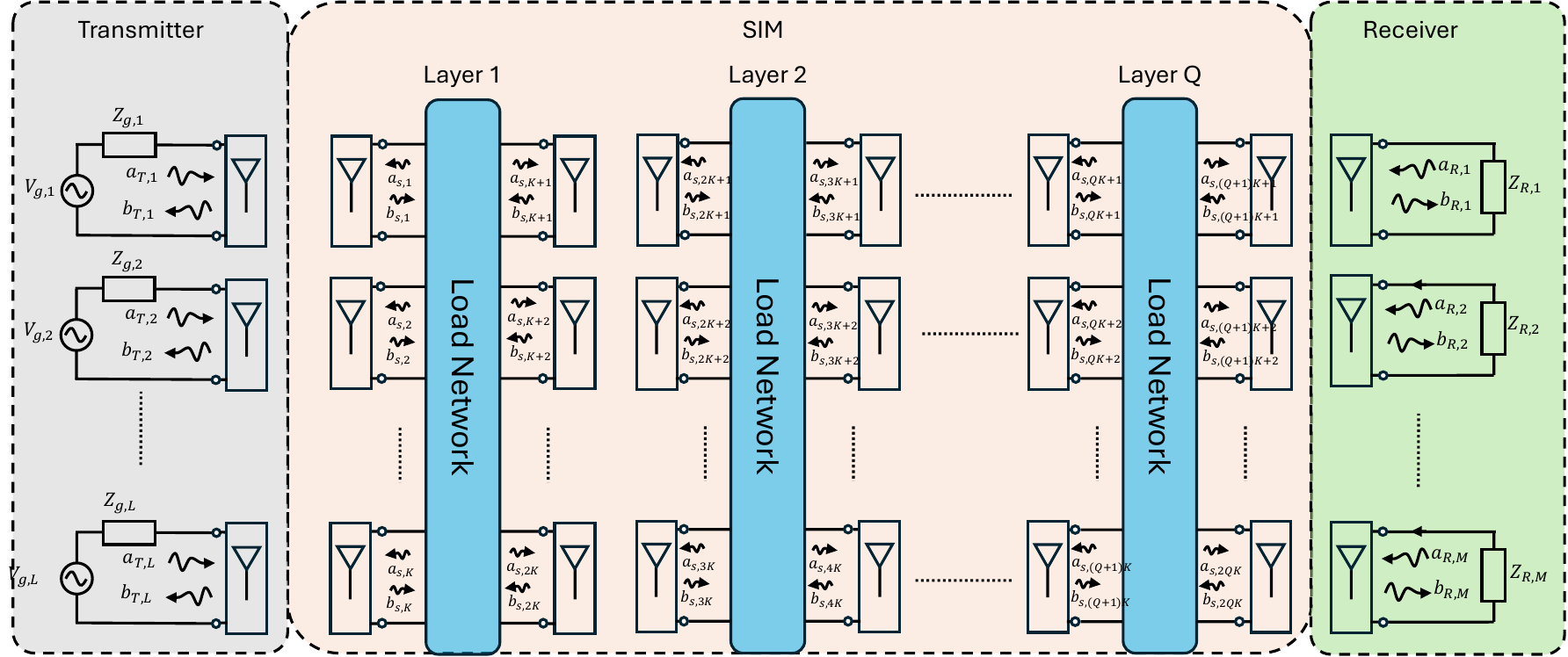}
	\caption{Multi-port SIM configuration.}
	\label{fig:simnetwork}
\end{figure}

Let us consider the multi-port SIM configuration shown in Fig.~\ref{fig:simnetwork}, which is characterized by a multi-antenna transmitter with $L$ ports, a multi-antenna receiver with $M$ ports, along with $N$ ports corresponding to the $N$ elements of a SIM that receives, processes, and retransmits electromagnetic waves.
Let us suppose the SIM composed of $Q$ stacked
T-RISs, referred in the following to as layers. Each layer is composed of a
receive array and a transmit array separated by one or more ground planes and
interconnected through tunable interconnection networks. Each array contains
$K$ antennas with accessible ports, so that each layer contributes $2K$ ports
to the overall network. Hence, the total number of SIM antenna ports is $N = 2QK$.
Following the multiport scattering formulation introduced in~\cite{Abrardo24RisOpt}, the overall electromagnetic structure can be described as
\begin{equation}
	\begin{bmatrix}
		\mathbf{b}_T\\
		\mathbf{b}_S\\
		\mathbf{b}_R
	\end{bmatrix}
	=
	\begin{bmatrix}
		\mathbf{S}_{TT} & \mathbf{S}_{TS} & \mathbf{S}_{TR}\\
		\mathbf{S}_{ST} & \mathbf{S}_{SS} & \mathbf{S}_{SR}\\
		\mathbf{S}_{RT} & \mathbf{S}_{RS} & \mathbf{S}_{RR}
	\end{bmatrix}
	\begin{bmatrix}
		\mathbf{a}_T\\
		\mathbf{a}_S\\
		\mathbf{a}_R
	\end{bmatrix},
	\label{eq:S_global}
\end{equation}
where $\mathbf{a}_x$ and $\mathbf{b}_x$, with $x\in\{T,S,R\}$, denote the vectors of incident and reflected waves at the transmitter, SIM, and receiver ports, respectively.

Assuming, as in \cite{Abrardo24RisOpt}, that both the transmitter and the receiver are matched to the reference impedance $Z_0$, we have $\mathbf{a}_R=\mathbf{0}$ and $\mathbf{a}_T=\mathbf{a}_g$.

The waves at the SIM ports satisfy the relationship:
\begin{equation}
	\mathbf{b}_S =
	\mathbf{S}_{ST}\mathbf{a}_T +
	\mathbf{S}_{SS}\mathbf{a}_S,
\end{equation}
and since the internal network imposes $\mathbf{a}_S = \mathbf{\Gamma}(\boldsymbol{\eta}) \mathbf{b}_S$, we obtain
\begin{equation}
	\mathbf{b}_S =
	\mathbf{S}_{ST}\mathbf{a}_T +
	\mathbf{S}_{SS}\mathbf{\Gamma}(\boldsymbol{\eta}) \mathbf{b}_S.
	\label{eq:bs_1}
\end{equation}
Therefore,
\begin{equation}
	\mathbf{b}_S =
	\left[\mathbf{I} - \mathbf{S}_{SS}\mathbf{\Gamma}(\boldsymbol{\eta})\right]^{-1}
	\mathbf{S}_{ST}\mathbf{a}_T.
	\label{eq:bS_solution}
\end{equation}
The received signal $\mathbf{b}_R $ can be written as
\begin{equation}
	\mathbf{b}_R =
	\mathbf{S}_{RT}\mathbf{a}_T +
	\mathbf{S}_{RS}\mathbf{a}_S
	=
	\mathbf{S}_{RT}\mathbf{a}_T +
	\mathbf{S}_{RS}\mathbf{\Gamma}(\boldsymbol{\eta})\mathbf{b}_S.
\end{equation}

By substituting \eqref{eq:bS_solution}, we obtain
\begin{equation}
	\mathbf{b}_R =
	\left[
	\mathbf{S}_{RT} +
	\mathbf{S}_{RS}\mathbf{\Gamma}(\boldsymbol{\eta})
	\mathbf{T}_S(\boldsymbol{\eta})
	\mathbf{S}_{ST}
	\right]\mathbf{a}_T.
\end{equation}
with the term
\begin{equation}
	\mathbf{T}_S(\boldsymbol{\eta}) =\mathbf{\Gamma}(\boldsymbol{\eta})
	\left[\mathbf{I} - \mathbf{S}_{SS}\mathbf{\Gamma}(\boldsymbol{\eta})\right]^{-1},
	\label{TS_def}
\end{equation}
representing the internal propagation operator of the SIM. 
Accordingly, the end-to-end transfer function of the SIM can be written as
\begin{equation}
	\mathbf{H}_{e2e}^{(S)}(\boldsymbol{\eta}) =
	\mathbf{S}_{RT}+\mathbf{S}_{RS}\mathbf{T}_S(\boldsymbol{\eta})\mathbf{S}_{ST}.
	\label{HS_final}
\end{equation}

The matrix $\mathbf{T}_S(\boldsymbol{\eta})$ represents the multiple internal interactions between the electromagnetic coupling among the SIM antennas and the active interconnection networks. This matrix is the scattering-domain counterpart of the impedance-domain matrix $[\mathbf{Z}_{SS}+\mathbf{Z}_S(\boldsymbol{\eta})]^{-1}$ used in the impedance formulation of the SIM model.

\subsection{Layered-isolated SIM model}
In this section, we present a simplified SIM model in the S-parameter domain,
obtained by reformulating the impedance-based model in~\cite{AbrardoSIM2025}.
The same physical assumptions adopted here have been also presented in  \cite{pettanice2026}, and the detailed discussion
is therefore omitted for the sake of brevity.

\begin{figure}[h!]
	\centering
	\includegraphics[width=\columnwidth]{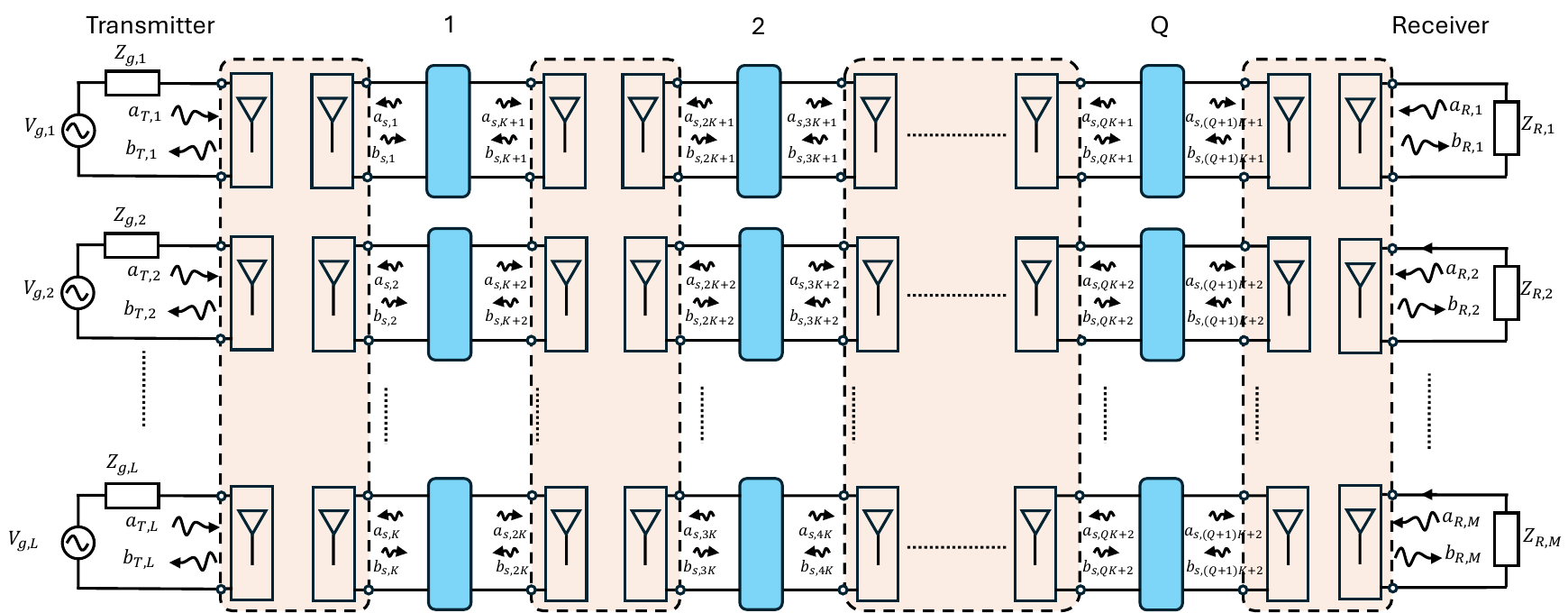}
	\caption{Multi-port SIM configuration composed of stacked T-RIS layers with one-to-one receive-transmit interconnection and electromagnetic isolation between adjacent layers.}
	\label{fig:simisolated}
\end{figure}

In this work we consider a one-to-one interconnection scheme, in which each
receive antenna is connected only to the corresponding transmit antenna of the
same element through an independent tunable two-port network, as shown in Fig. \ref{fig:simisolated}. 

Each
receive–transmit pair together with the corresponding two-port network forms a
unit cell. Under this assumption, the overall SIM interconnection network is
composed of $QK$ independent tunable two-port cells, and the corresponding
multiport interconnection matrix has a block-diagonal (diagonal) structure.

The SIM antenna ports are indexed layer-wise. The ports of the $q$-th layer
occupy the index range $\{2K(q-1)+1,\ldots,2Kq\}$. Within each layer, the
$k$-th unit cell connects the two ports with local indices $k$ and $k+K$.
The corresponding global port indices are denoted by $m(q,k)$ and $n(q,k)$.
Further implementation details and examples of physical realizations can be
found in \cite{pettanice2026}.

Let us now assume that:
\begin{enumerate}
	\item 
	the multi-antenna transmitter is electromagnetically coupled only with the 
	receive array of the first SIM layer, and the multi-antenna receiver is
	electromagnetically coupled only with the transmit array of the last SIM
	layer. This means that the coupling between the transmitter/receiver and the
	internal SIM layers is negligible.
	\item 
	the receive and transmit arrays composing each T-RIS layer are not
	directly coupled except through the tunable interconnection network.
	\item 
	across the SIM, electromagnetic coupling occurs only between adjacent
	arrays through propagation in the surrounding medium.
\end{enumerate}

Although idealized, these assumptions are physically meaningful and
can be closely approached in practice through suitable design strategies.
Both the isolation between the receive and transmit arrays within the
same layer (Assumption~2) and the suppression of coupling between
non-adjacent layers (Assumption~3) are enforced when the multi-layer
structure is enclosed in a metallic housing lined with microwave
absorbing materials: in this case, isolation is guaranteed by the
geometry of the enclosure itself, independently of the finite size of
the ground planes. The unidirectionality of the integrated amplifiers
further suppresses inter-layer coupling by preventing backward wave
propagation~\cite{Liu2022}. A quantitative assessment of the isolation
levels achievable with this approach is provided in
Appendix~\ref{app:isolation}.

In the full-wave simulations of Section~\ref{sec:results}, infinite
ground planes are assumed in FEKO as a computationally efficient
surrogate for the absorber-lined enclosure, enforcing both isolation
conditions exactly while avoiding the overhead of a full absorber model.

Under the above assumptions, the SIM exhibits a structured electromagnetic
coupling pattern in which only neighboring arrays interact directly. As
illustrated in Fig.~\ref{fig:simisolated}, the transmitter couples only to
the first SIM layer, the receiver only to the last layer, and interactions
inside the SIM occur only between adjacent layers.

From a network-theoretic perspective, the electromagnetic coupling among all
SIM antenna ports is modeled as a linear multiport scattering network described
by the matrix $\mathbf{S}_{SS}\in\mathbb{C}^{N\times N}$, which depends only on the
geometry of the structure and the propagation environment. 
Therefore, similarly to the impedance-based formulation, it is independent of the configuration of the internal load/interconnection networks and can be obtained by full-wave electromagnetic simulations.
The assumptions (1)-(3) induce a structured sparsity in the internal scattering
matrix $\mathbf{S}_{SS}$, which becomes block diagonal (or block-banded with
nearest-neighbor interactions) when the ports are ordered according to the SIM
layers. This structure is fully analogous to the impedance-matrix structure
obtained in~\cite{AbrardoSIM2025}. Consequently, the same algebraic framework and
computational strategy derived in~\cite{AbrardoSIM2025} can be directly applied in
the S-parameter formulation.

Under these assumptions, the matrix $\mathbf{S}_{SS}$ can be conveniently
represented in block form as
\begin{equation}
	\small
	\mathbf{S}_{SS} =
	\begin{bmatrix}
		\mathbf{\tilde{S}}^{(0)}_{2,2} & \mathbf{0} & \mathbf{0} & \mathbf{0} & \cdots & \mathbf{0} \\
		\mathbf{0} & \mathbf{\tilde{S}}^{(1)}_{1,1} & \mathbf{\tilde{S}}^{(1)}_{1,2} & \mathbf{0} & \cdots & \mathbf{0}\\
		\mathbf{0} & \mathbf{\tilde{S}}^{(1)}_{2,1} & \mathbf{\tilde{S}}^{(1)}_{2,2} & \mathbf{0} & \cdots & \mathbf{0}\\
		\mathbf{0} & \mathbf{0} & \mathbf{0} & \mathbf{\tilde{S}}^{(2)}_{1,1} & \mathbf{\tilde{S}}^{(2)}_{1,2} & \cdots \\
		\mathbf{0} & \mathbf{0} & \mathbf{0} & \mathbf{\tilde{S}}^{(2)}_{2,1} & \mathbf{\tilde{S}}^{(2)}_{2,2} & \cdots \\
		\vdots & \vdots & \vdots & \vdots & \vdots & \ddots \\
		\mathbf{0} & \mathbf{0} & \mathbf{0} & \mathbf{0} & \cdots & \mathbf{\tilde{S}}^{(Q)}_{1,1}
	\end{bmatrix}
	\label{S_SS}
\end{equation}
where $\mathbf{\tilde{S}}^{(u)}_{i,j} \in \mathbb{C}^{K \times K}$, with
$u\in\{0,\ldots,Q\}$, denote the electromagnetic coupling sub-matrices
associated with free-space propagation between two consecutive arrays in the
stacked structure. 
In particular, the case $u=0$ corresponds to the single block
$\mathbf{\tilde{S}}^{(0)}_{2,2}$, which describes the mutual coupling
among the antenna elements of the receive array of the first layer.
Similarly, the case $u=Q$ corresponds to $\mathbf{\tilde{S}}^{(Q)}_{1,1}$,
which describes the mutual coupling within the transmit array of the last
layer. For each intermediate case $u\in\{1,\ldots,Q-1\}$, 
the subindex $i=1$ refers to the transmit array of layer $q$ and $i=2$
to the receive array of layer $q+1$; the diagonal blocks
$\mathbf{\tilde{S}}^{(u)}_{i,i}$, $i\in\{1,2\}$, capture the mutual
coupling among the antenna elements within each array, while the
off-diagonal blocks $\mathbf{\tilde{S}}^{(u)}_{i,j}$, $i\neq j$,
describe the electromagnetic coupling between the transmit array of
layer $q$ and the receive array of layer $q+1$.
These matrices depend
only on the electromagnetic propagation between the arrays
and can be computed using full-wave simulations.

\subsection{Unit-cell tunable two-port network model}
In order to complete the SIM model, we now introduce the model of the internal
interconnection network, which defines the relation between the waves at the
SIM ports and leads to the definition of the internal propagation operator
$\mathbf{T}_S(\boldsymbol{\eta})$.
As mentioned in the above subsection, each unit cell consists of a receive antenna and a transmit antenna internally connected through a tunable non-reciprocal two-port network. Here, the two-port
network is modeled as an ideal matched unilateral device with controllable
phase shift and constant gain.
In particular, we consider a two-port network implemented as the cascade of an
ideal unilateral transistor amplifier and a reactive phase-shift network.
Under matched conditions, the scattering matrix of the $k$-th unit cell of the
$q$-th layer can be written as
\begin{equation}
	\mathbf{\Gamma}^{(q)}_k =
	\begin{bmatrix}
		0 & 0\\
		G e^{j\eta_{q,k}} & 0
	\end{bmatrix},
	\label{eq:cell_S}
\end{equation}
where $G$ denotes the power gain (or attenuation factor) and $\eta_{q,k}$ is
the tunable phase shift associated with the $(q,k)$-th unit cell. This model
represents an ideal unilateral device that allows signal transmission only
from the receive antenna to the transmit antenna, while isolating the reverse
direction.

By collecting the $K$ unit cells of the $q$-th layer, we define the diagonal
matrix
\begin{equation}
	\mathbf{G}^{(q)} =
	G\,\mathrm{diag}\!\left(e^{j\boldsymbol{\eta}_q}\right)
	\in \mathbb{C}^{K \times K},
    \label{eq:Gdiag}
\end{equation}
where $\boldsymbol{\eta}_q = [\eta_{q,1},\ldots,\eta_{q,K}]^T$.

With a suitable ordering of the SIM ports, the global scattering matrix of the
SIM internal interconnection network can be written as a block matrix
\begin{equation}
	\small
	\mathbf{\Gamma}(\boldsymbol{\eta}) =
	\begin{bmatrix}
		\mathbf{0} & \mathbf{0} & \mathbf{0} & \cdots & \mathbf{0} \\
		\mathbf{G}^{(1)} & \mathbf{0} & \mathbf{0} & \cdots & \mathbf{0} \\
		\mathbf{0} & \mathbf{0} & \mathbf{0} & \cdots & \mathbf{0} \\
		\mathbf{0} & \mathbf{0} & \mathbf{G}^{(2)} & \cdots & \mathbf{0} \\
		\vdots & \vdots & \vdots & \ddots & \vdots \\
		\mathbf{0} & \mathbf{0} & \mathbf{0} & \cdots & \mathbf{G}^{(Q)} & \mathbf{0}
	\end{bmatrix},
	\label{Gamma_SIM}
\end{equation}
which is block diagonal with $Q$ non-zero submatrices, each corresponding to
one SIM layer. 

It is worth emphasizing that the matrix $\mathbf{\Gamma}(\boldsymbol{\eta})$
models only the internal interconnection networks of the SIM, whereas the
matrix $\mathbf{S}_{SS}$ models the electromagnetic propagation between the
antenna arrays. The overall SIM behavior is therefore determined by the
interaction between the propagation matrix $\mathbf{S}_{SS}$ and the internal
interconnection matrix $\mathbf{\Gamma}(\boldsymbol{\eta})$.
Since the interconnection network includes active unilateral devices, the SIM
is not a passive structure and introduces a feed-forward propagation mechanism across the layers.
The structured modeling of signal propagation within the SIM is addressed in the next section.

The unilateral two-port network model adopted in this paper is directly
inspired by the SCS architecture of~\cite{Taravati2017,Ma2019,Liu2022}.
The gain values considered ($G \leq 6$~dB) are moderate and compatible
with standard RF amplifier technologies (e.g., CMOS or GaAs). The total
power consumption scales linearly with the number of active unit cells
$QK$, and the effect of amplifier noise on system performance is
discussed in Section~\ref{sec:results}.

\subsection{Stability of the Active SIM}

The end-to-end transfer function derived in \eqref{HS_final} involves
the operator $(\mathbf{I} - \mathbf{S}_{SS}\mathbf{\Gamma}(\boldsymbol{\eta}))^{-1}$,
which plays a central role in the physical interpretation of signal
propagation within the SIM.
To make this explicit, we expand this operator as a Neumann series:
\begin{equation}
\left(\mathbf{I} - \mathbf{S}_{SS}\mathbf{\Gamma}\right)^{-1}
=
\mathbf{I}
+ \mathbf{S}_{SS}\mathbf{\Gamma}
+ \left(\mathbf{S}_{SS}\mathbf{\Gamma}\right)^2
+ \left(\mathbf{S}_{SS}\mathbf{\Gamma}\right)^3
+ \cdots
\label{eq:neumann}
\end{equation}
Each term in \eqref{eq:neumann} has a clear physical meaning.
The matrix $\mathbf{\Gamma}(\boldsymbol{\eta})$ describes one pass
through the active interconnection networks, while $\mathbf{S}_{SS}$
describes the electromagnetic coupling among the SIM antenna ports.
Therefore, the product $\mathbf{S}_{SS}\mathbf{\Gamma}(\boldsymbol{\eta})$
represents one complete internal round-trip: the signal is first
processed by the active networks and then coupled back through the
antenna structure.
The Neumann series sums all contributions corresponding to zero,
one, two, and more round-trips, and is therefore the classical
representation of a feedback operator.

The series in \eqref{eq:neumann} converges if and only if the
spectral radius of the round-trip operator satisfies
\begin{equation}
\rho\left(\mathbf{S}_{SS}\mathbf{\Gamma}(\boldsymbol{\eta})\right) < 1,
\label{eq:stability}
\end{equation}
where $\rho(\mathbf{A}) = \max_i |\lambda_i(\mathbf{A})|$ denotes
the spectral radius of a matrix, i.e., the largest absolute value
among all its eigenvalues.
Physically, $\rho(\mathbf{S}_{SS}\mathbf{\Gamma})$ measures the
maximum amplification factor that the signal can experience after
one internal round-trip, over all possible excitation patterns.
When $\rho < 1$, every round-trip attenuates the signal in all
directions, the series \eqref{eq:neumann} converges, and the
system is stable.
When $\rho \geq 1$, there exists at least one excitation pattern
for which the signal is not attenuated after each round-trip,
the series diverges, and the system can exhibit oscillatory or
unbounded behavior.

Under the layered-isolated assumptions of
Section~\ref{sec:SIM_model} and the unilateral active two-port
model in \eqref{eq:cell_S}, the matrix
$\mathbf{\Gamma}(\boldsymbol{\eta})$ is strictly
block-lower-triangular (see \eqref{Gamma_SIM}).
Combined with the block-banded structure of $\mathbf{S}_{SS}$
(see \eqref{S_SS}), the product
$\mathbf{S}_{SS}\mathbf{\Gamma}(\boldsymbol{\eta})$ is nilpotent,
meaning that $\left(\mathbf{S}_{SS}\mathbf{\Gamma}\right)^N = \mathbf{0}$
for a finite integer $N$.
A nilpotent matrix has all eigenvalues equal to zero, so
$\rho\left(\mathbf{S}_{SS}\mathbf{\Gamma}\right) = 0$, and
condition \eqref{eq:stability} is always satisfied.
As a consequence, the Neumann series \eqref{eq:neumann} terminates
after a finite number of terms, a result that will find a direct
counterpart in the recursive feed-forward representation to be derived in
Section~\ref{sec:recursive}.
The unilateral active SIM is therefore \emph{unconditionally stable},
regardless of the gain value $G$ and the phase configuration
$\boldsymbol{\eta}$.
This is a structural advantage of the unilateral architecture:
unlike bilateral or reciprocal active designs, stability does not
need to be verified case by case but is guaranteed by the
feed-forward structure itself.

\section{Recursive end-to-end transfer function}
\label{sec:recursive}
In this section, we derive a feed-forward transfer function by directly exploiting the layered
organization of the SIM. This avoids the explicit inversion of the global
multi-port operator and leads to a recursive description that is consistent
with the electromagnetic architecture of the structure.

To make the recursive structure explicit, let us order the SIM ports layer-wise as 
\begin{equation}
\big[\mathbf{a}_{r}^{(1)},\mathbf{a}_{t}^{(1)},
\mathbf{a}_{r}^{(2)},\mathbf{a}_{t}^{(2)},\ldots,
\mathbf{a}_{r}^{(Q)},\mathbf{a}_{t}^{(Q)}\big],
\label{eq:layer-wise}
\end{equation}
where $\mathbf{a}_{r}^{(q)}\in\mathbb{C}^{K\times 1}$ and $\mathbf{a}_{t}^{(q)}\in\mathbb{C}^{K\times 1}$ collect the incident
waves at the receive and transmit ports of the $q$-th layer, respectively.

Under the unilateral active two-port model, the relation between the local
receive and transmit ports of layer $q$ is
\begin{equation}
	\mathbf{a}_{t}^{(q)}
	=
	\mathbf{G}^{(q)}\mathbf{b}_{r}^{(q)}
	\qquad q=1,\ldots,Q,
	\label{eq:local_unilateral}
\end{equation}
where $\mathbf{G}^{(q)}$ is defined in \eqref{eq:Gdiag}. 

Due to the nearest-neighbor interaction pattern described by
\eqref{S_SS}, the only non-zero inter-layer coupling from the transmit ports of
layer \(q\) to the receive ports of layer \(q+1\) is given by the block
\[
\mathbf{\tilde{S}}_{2,1}^{(q)} \in \mathbb{C}^{K\times K}.
\]
Hence, the SIM can be viewed as a cascade of local diagonal operators
\(\mathbf{G}^{(q)}\) and nearest-neighbor electromagnetic coupling blocks
\(\mathbf{\tilde{S}}_{2,1}^{(q)}\).

To make the structured representation consistent with the multi-port
formulation, we explicitly introduce the operators that describe the coupling
between the external arrays and the SIM.

Let us denote by $\mathbf{H}_{TS} \in \mathbb{C}^{K \times L}$ the transfer
matrix from the transmitter to the receive side of the first SIM layer, and by
$\mathbf{H}_{SR} \in \mathbb{C}^{M \times K}$ the transfer matrix from the
transmit side of the last SIM layer to the receiver.
In the absence of additional processing blocks (e.g., filtering or combining
operations at the transmitter or receiver), these matrices coincide with the
corresponding electromagnetic coupling sub-matrices in the S-parameter representation \eqref{eq:S_global}.

More precisely, these matrices are obtained as suitable sub-matrices of the global scattering
matrices $\mathbf{S}_{ST}$ and $\mathbf{S}_{RS}$, respectively. In particular,
by ordering the SIM ports layer-wise as in \eqref{eq:layer-wise}, we define
\begin{equation}
	\mathbf{H}_{TS} = \mathbf{S}_{ST}^{(r,1)}, \qquad
	\mathbf{H}_{SR} = \mathbf{S}_{RS}^{(t,Q)},
\end{equation}
where $\mathbf{S}_{ST}^{(r,1)}$ denotes the sub-matrix of $\mathbf{S}_{ST}$
obtained by selecting the rows corresponding to the receive ports of the first
SIM layer, and $\mathbf{S}_{RS}^{(t,Q)}$ denotes the sub-matrix of
$\mathbf{S}_{RS}$ obtained by selecting the columns corresponding to the
transmit ports of the last SIM layer.




Due to the unilateral interconnection and the nearest-neighbor coupling structure, the field propagates layer by layer according to the recursion
\begin{align}
	\mathbf{T}_c^{(1)} &= \mathbf{H}_{TS},
	\label{eq:Tc1_fwd}\\
	\mathbf{T}_c^{(q)} &=
	\mathbf{\tilde{S}}_{2,1}^{(q-1)}
	\mathbf{G}^{(q-1)}
	\mathbf{T}_c^{(q-1)},
	\qquad q=2,\ldots,Q.
	\label{eq:Tc_rec_fwd}
\end{align}
Here, \(\mathbf{T}_c^{(q)}\) represents the signal transfer from the transmitter
to the receive ports of layer \(q\).

The useful transfer from the multi-antenna transmitter to the transmit ports of the last SIM
layer is then
\begin{equation}
	\mathbf{H}_2(\boldsymbol{\eta})
	=
	\mathbf{G}^{(Q)}\mathbf{T}_c^{(Q)}.
	\label{eq:H2_struct}
\end{equation}

Taking into account the definition of $\mathbf{H}_{SR}$, and assuming that the direct coupling between the transmitter and the receiver is negligible (i.e., $\mathbf{S}_{RT} \approx \mathbf{0}$), the structured end-to-end transfer matrix reduces to

\begin{equation}
	\mathbf{H}_{e2e}^{(S)}(\boldsymbol{\eta})
	= \mathbf{H}_{SR}\,\mathbf{H}_2(\boldsymbol{\eta}).
	\label{eq:He2e_struct_new}
\end{equation}
where $\mathbf{H}_2(\boldsymbol{\eta})$ describes the propagation through the
SIM layers from the receive side of the first layer to the transmit side of
the last layer.
Equation \eqref{eq:He2e_struct_new} provides the desired end-to-end transfer
without requiring the inversion of the global multi-port matrix.
This recursive formulation is algebraically equivalent to the global
multi-port solution in \eqref{HS_final}, but avoids the explicit inversion
of the matrix $[\mathbf{I} - \mathbf{S}_{SS}\mathbf{\Gamma}(\boldsymbol{\eta})]$.


\section{SIM optimization}\label{SIM_opt}

The SIM configuration problem is formulated as the optimization of a
physics-based input-output mapping. Given a finite set of input excitations
and corresponding desired output responses, the SIM control vector
\(\boldsymbol{\eta}\) is determined by minimizing a suitable loss function.

Let the \(I\) input excitations be collected in the source matrix
\[
\mathbf{A}_T=\big[\mathbf{a}_{T,1},\ldots,\mathbf{a}_{T,I}\big],
\]
and let
\[
\mathbf{A}_S=\big[\mathbf{a}_{S,1},\ldots,\mathbf{a}_{S,I}\big], \qquad
\mathbf{B}_S=\big[\mathbf{b}_{S,1},\ldots,\mathbf{b}_{S,I}\big],
\]
denote the corresponding incident and reflected SIM wave matrices. The matrix
collecting the SIM output responses is defined as
\[
\mathbf{\hat{X}}(\boldsymbol{\eta})
=
\big[\mathbf{\hat{x}}_1(\boldsymbol{\eta}),\ldots,\mathbf{\hat{x}}_I(\boldsymbol{\eta})\big].
\]

In the noiseless setting, the relevant quantities satisfy
\begin{align}
\mathbf{B}_S &= \mathbf{S}_{ST}\mathbf{A}_T + \mathbf{S}_{SS}\mathbf{A}_S,
\label{eq:BS_mat_new}\\
\mathbf{\hat{X}}(\boldsymbol{\eta}) &= \mathbf{S}_{RT}\mathbf{A}_T + \mathbf{S}_{RS}\mathbf{A}_S.
\label{eq:Xhat_mat_new}
\end{align}

Given a desired output matrix \(\mathbf{X}_d\in\mathbb{C}^{M\times I}\), we
introduce the error matrix
\[
\mathbf{E}(\boldsymbol{\eta})
:=
\beta\,\mathbf{\hat{X}}(\boldsymbol{\eta}) - \mathbf{X}_d,
\]
where \(\beta\in\mathbb{C}\) is a complex scaling factor. The corresponding loss
function is
\begin{equation}
L(\boldsymbol{\eta}) = \|\mathbf{E}(\boldsymbol{\eta})\|_F^2.
\label{eq:loss_new}
\end{equation}

The SIM optimization problem is therefore
\[
\min_{\boldsymbol{\eta}} \; L(\boldsymbol{\eta}).
\]

As shown in Section~\ref{sec:SIM_model}, thanks to the layered structure of the
SIM, the useful transfer matrix from the transmitter to the transmit side of
the last SIM layer can be evaluated recursively by using \eqref{eq:H2_struct}
where the forward factors $\mathbf{T}_c^{(Q)}$ are obtained from the
layer-by-layer recursion introduced in \eqref{eq:Tc1_fwd}--\eqref{eq:Tc_rec_fwd}.

Accordingly, the SIM output matrix can be written as
\begin{equation}
\mathbf{\hat{X}}(\boldsymbol{\eta})
=\mathbf{A}\,\mathbf{H}_2(\boldsymbol{\eta})\,\mathbf{\Lambda},
\label{eq:Xhat_fast}
\end{equation}
where \(\mathbf{A}\), and \(\mathbf{\Lambda}\) collect the fixed
contributions associated with the receiver-side coupling, and
the input excitations, respectively.

For a given \(\boldsymbol{\eta}\), the optimal complex scaling factor is
obtained in closed form as
\begin{equation}
\beta^\star
=
\frac{\mathrm{trace}\!\left(
\big(\mathbf{A}\mathbf{H}_2\mathbf{\Lambda}\big)
\mathbf{X}_d^H
\right)}
{\mathrm{trace}\!\left(
\big(\mathbf{A}\mathbf{H}_2\mathbf{\Lambda}\big)
\big(\mathbf{A}\mathbf{H}_2\mathbf{\Lambda}\big)^H
\right)}.
\label{eq:beta_opt_struct}
\end{equation}

\paragraph{Structured gradient evaluation.}
To evaluate the gradient efficiently, 
we first define the backward transfer matrices
\(\mathbf{T}_r^{(q)}\), which map a perturbation generated at the transmit ports
of layer \(q\) to the transmit ports of the last SIM layer. Analogously to \eqref{eq:Tc1_fwd}-\eqref{eq:Tc_rec_fwd}, these matrices are
recursively defined as
\begin{align}
	\mathbf{T}_r^{(Q)} &= \mathbf{I},
	\label{eq:TrQ_bwd}\\
	\mathbf{T}_r^{(q)} &=
	\mathbf{T}_r^{(q+1)}
	\mathbf{G}^{(q+1)}
	\mathbf{\tilde{S}}_{2,1}^{(q)},
	\qquad q=Q-1,\ldots,1.
	\label{eq:Tr_rec_bwd}
\end{align}
Indeed, for $q<Q$, a perturbation generated at layer $q$ must first pass
through the coupling block $\mathbf{\tilde{S}}_{2,1}^{(q)}$, then through the
local active operator $\mathbf{G}^{(q+1)}$, and so on until the last layer.
Then, let $\eta_p$ denote the control parameter associated with the $k$-th unit cell
of layer \(q\), and let \(\mathbf{G}^{(q)}\) be the corresponding diagonal
local operator. Since only one diagonal entry of \(\mathbf{G}^{(q)}\) depends
on \(\eta_p\), its derivative
\[
\frac{\partial \mathbf{G}^{(q)}}{\partial \eta_p}
\]
has a single non-zero entry.

Using the previously defined forward and backward transfer matrices
\(\mathbf{T}_c^{(q)}\) and \(\mathbf{T}_r^{(q)}\), the derivative of the useful
transfer matrix admits the factorized form
\begin{equation}
\frac{\partial \mathbf{H}_2}{\partial \eta_p}
=
\mathbf{T}_r^{(q)}
\frac{\partial \mathbf{G}^{(q)}}{\partial \eta_p}
\mathbf{T}_c^{(q)}.
\label{eq:dH2_sep_new}
\end{equation}

Let
\[
\mathbf{U}(\boldsymbol{\eta}) := \beta\,\mathbf{A}\mathbf{H}_2(\boldsymbol{\eta})\mathbf{\Lambda}.
\]
Then
\begin{equation}
\frac{\partial \mathbf{U}}{\partial \eta_p}
=
\beta\,
\mathbf{A}\,\mathbf{T}_r^{(q)}
\frac{\partial \mathbf{G}^{(q)}}{\partial \eta_p}
\mathbf{T}_c^{(q)}\,\mathbf{\Lambda}.
\end{equation}
Defining
\[
\mathbf{A}_q := \mathbf{A}\mathbf{T}_r^{(q)},
\qquad
\mathbf{B}_q := \mathbf{T}_c^{(q)}\mathbf{\Lambda},
\]
one obtains the fully factorized expression
\begin{equation}
\frac{\partial \mathbf{U}}{\partial \eta_p}
=
\beta\,
\mathbf{A}_q
\frac{\partial \mathbf{G}^{(q)}}{\partial \eta_p}
\mathbf{B}_q.
\end{equation}

Finally, the gradient of the loss is computed as
\begin{equation}
\frac{\partial L}{\partial \eta_p}
=
2\,\Re\!\left\{
\sum_{i=1}^I
\mathbf{r}_i^H
\frac{\partial \mathbf{U}_i}{\partial \eta_p}
\right\},
\label{eq:grad_struct_final}
\end{equation}
where \(\mathbf{r}_i\) denotes the residual associated with the \(i\)-th
training excitation.

\paragraph{Gradient-descent algorithm.}
The minimization of \eqref{eq:loss_new} is carried out through a gradient-descent
iteration combined with a backtracking line search satisfying the Armijo
sufficient-decrease condition.

Let \(\boldsymbol{\eta}^{(t)}\) denote the SIM control vector at iteration \(t\).
Starting from an initial point \(\boldsymbol{\eta}^{(0)}\), the update is
\begin{equation}
\boldsymbol{\eta}^{(t+1)}
=
\boldsymbol{\eta}^{(t)}
-
\alpha^{(t)}
\nabla_{\boldsymbol{\eta}}L\!\left(\boldsymbol{\eta}^{(t)}\right),
\label{eq:gd_update_new}
\end{equation}
where \(\alpha^{(t)}>0\) is the step size. Since the control variables are
unconstrained real-valued phases, no projection step is required.

The step size is initialized at \(\alpha^{(t)}=1\) and is iteratively reduced
according to
\[
\alpha^{(t)} \leftarrow \beta_{\mathrm{BT}}\alpha^{(t)},
\qquad 0<\beta_{\mathrm{BT}}<1,
\]
until the Armijo condition
\begin{equation}
L\!\left(\boldsymbol{\eta}^{(t+1)}\right)
\le
L\!\left(\boldsymbol{\eta}^{(t)}\right)
-
\frac{\alpha^{(t)}}{2}
\left\|
\nabla_{\boldsymbol{\eta}}L\!\left(\boldsymbol{\eta}^{(t)}\right)
\right\|_2^2
\label{eq:armijo_new}
\end{equation}
is satisfied.
Note that although the proposed gradient-based approach already provides a favorable trade-off between convergence behavior and computational complexity as shown in \cite{AbrardoSIM2025}, alternative optimization strategies could also be considered. In particular, adaptive gradient methods, quasi-Newton schemes, or stochastic optimization techniques may further accelerate convergence in strongly coupled SIM configurations. A systematic investigation of such approaches is left for future work.

\paragraph{Per-iteration complexity.}

Let \(Q\) denote the number of SIM layers and \(K\) the number of unit cells per layer. As shown in \cite{AbrardoSIM2025}, for a general electromagnetic collaborative object (ECO), i.e., an electromagnetic structure not organized in a layered SIM architecture, the computational complexity of each optimization step scales as \(\mathcal{O}((QK)^3)\), due to the manipulation of a large dense matrix. This rapidly becomes computationally prohibitive for electrically large structures. By exploiting the layered organization of a conventional SIM, it was shown in \cite{AbrardoSIM2025} that the complexity can be reduced to \(\mathcal{O}(QK^3)\), even in the presence of reciprocal passive interconnection networks. This reduction originates from the block-banded structure induced by the nearest-neighbor interaction model. In the present work, the introduction of unilateral active two-port networks yields an additional structural simplification. Specifically, the feed-forward propagation mechanism induced by the unilateral architecture allows the end-to-end transfer function to be computed through simple forward and backward recursions, avoiding the matrix inversions required by conventional reciprocal SIM formulations. As a result, the matrices \(\mathbf{T}_c^{(q)}\) and \(\mathbf{T}_r^{(q)}\) are obtained through products of \(K\times K\) matrices, and the dominant structural cost per iteration scales as \(\mathcal{O}(QK^2)\). Once these quantities are available, the loss function and its gradient can be evaluated through simple accumulations over the training samples. Note that, in the considered design problem, the training set does not correspond to a large collection of input-output examples as in conventional machine learning applications. Instead, it consists of a small number of linearly independent excitation vectors defining the desired end-to-end transformation. For the $N_s \times N_s$ MIMO scenarios considered in this work, a natural choice is to use the canonical basis vectors and the corresponding desired outputs, yielding $I=N_s$ training samples. Consequently, $I$ remains small and does not scale with the SIM dimensions $Q$ and $K$. Therefore, the dependence on the number of training excitations only affects the final accumulation stage, whereas the dominant complexity reduction stems from the recursive feed-forward computation of the signal propagation through the SIM. Overall, the proposed unilateral active SIM architecture reduces the complexity from \(\mathcal{O}((QK)^3)\) for a general ECO, to \(\mathcal{O}(QK^3)\) for a conventional reciprocal layered SIM, and finally to \(\mathcal{O}(QK^2)\) for the proposed unilateral architecture, while preserving the electromagnetic accuracy of the underlying multi-port formulation.

\section{Numerical results}
\label{sec:results}
In this section, we present a set of representative results obtained by
considering different SIM-based functionalities. The goal is to highlight how
the proposed modeling based on active two-port networks not only provides a
significant computational advantage, thanks to the unilateral structure
discussed in the previous sections, but also enables a highly flexible
optimization of the electromagnetic response. At the same time, the use of active elements allows us to compensate for the
attenuation that would inherently arise in passive SIM configurations, even
when relatively small gains \(G\) are employed.

More specifically, in the following discussion we focus on the mechanisms that
govern the overall signal attenuation through the SIM. In particular, the
end-to-end performance is mainly determined by the interplay among the following key
factors.

The propagation of the useful signal through the SIM can be interpreted as a
cascade of stages, where each layer contributes both a beamforming gain and a
propagation loss. The latter is governed by the inter-layer coupling
coefficients \(\mathbf{S}_{2,1}\), which decrease as the distance between layers
increases. As a result, larger spacing between layers leads to stronger
attenuation of the signal as it propagates through the structure.

The active gain \(G\) plays a key role in compensating for this attenuation.
Even moderate values of \(G\) can effectively counterbalance the losses
introduced by the inter-layer propagation, thus preserving the signal strength
across multiple layers.

In addition, the number of elements \(K\) within each layer provides a coherent
beamforming gain that further enhances the useful signal. Unlike propagation
effects, this gain scales constructively with the array size, making larger
layers particularly beneficial in improving the overall performance.

To validate the above considerations, the performance of the optimized SIM configurations is assessed
through electromagnetic simulations, which provide an accurate representation
of the underlying physical system.

\subsection{Simulation Scenario}
\label{subsec:simulation_setup}

\begin{figure}
	\centering
	\includegraphics[width=1\columnwidth]{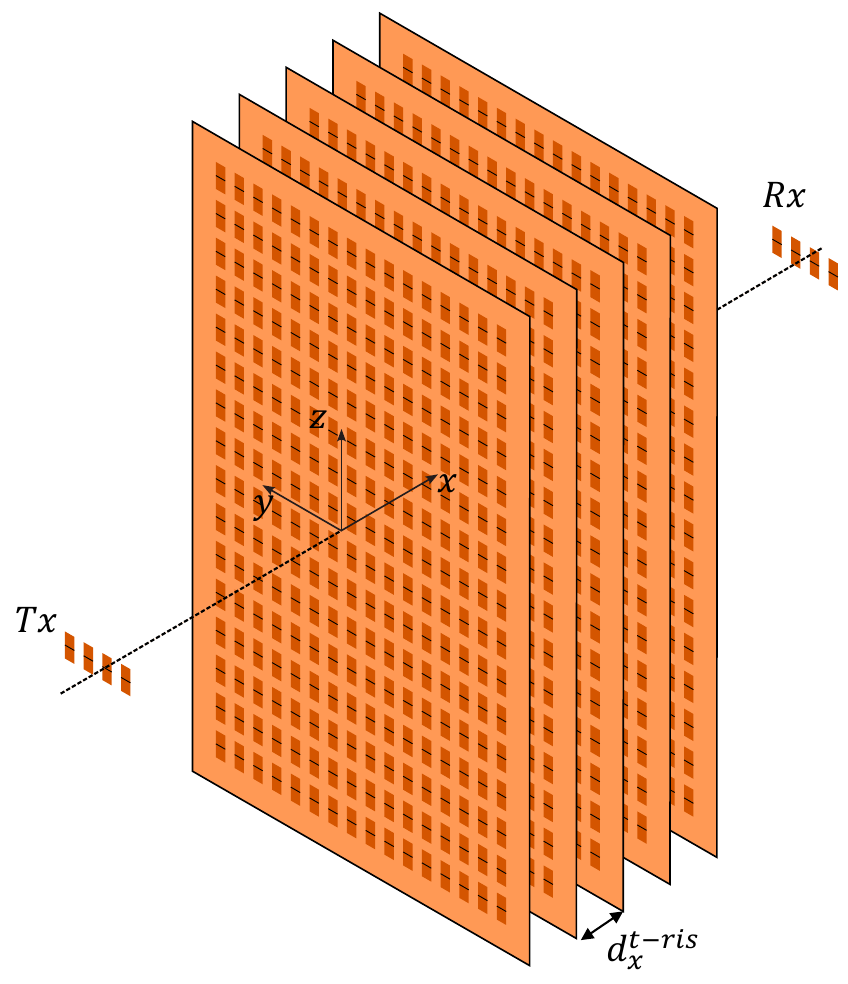}
	\caption{\textcolor{black}{Structure of the considered SIM with $Q$ stacked T-RIS layers. Each layer includes planar receive and transmit arrays interconnected by tunable active phase-shifting networks.}}
	\vspace{-2mm}
	\label{fig:SIMmodel}
\end{figure}
The general SIM architecture adopted in the following examples is depicted in Fig.~\ref{fig:SIMmodel}. The structure consists of $Q$ T-RIS layers arranged in a stacked configuration, parallel to each other and aligned along the $x$-axis of a global reference frame. The system operates at a carrier frequency $f_0 = 28$~GHz, corresponding to a free-space wavelength $\lambda_0 = 10.7\ \mathrm{mm}$.

Each element of the $q$-th T-RIS layer, with $q = 1,2,\ldots,Q$, is modeled as a center-fed planar metallic strip dipole with width $w_d = 0.05\lambda$ and length $l_d = 0.473\lambda$. The dipoles are oriented along the $y$ and $z$ directions and are positioned above a ground plane at a distance $h_d = 0.25\lambda$. Within each layer, the elements are arranged in a uniform planar array (UPA) with $N_y^{q}$ elements along the $y$-axis and $N_z^{q}$ elements along the $z$-axis. The inter-element spacing is set to $d_y^{dip} = \lambda/2$ along $y$ and $d_z^{dip} = (3/4)\lambda$ along $z$.

The separation between consecutive T-RIS layers, denoted by $d_x^{\,t\text{-}ris}$, is assumed to be the same for all layers. As illustrated in Fig.~\ref{fig:SIMmodel}, each layer is composed of a receive array and a transmit array including $K$ unit cells. Each unit cell consists of a receiving and a transmitting strip dipole located on opposite sides of the ground plane and interconnected through a linear active two-port network. This network enables the tuning of the phase response of each unit cell, while the gain $G$ is assumed to be fixed.

The overall electromagnetic behavior of the SIM is therefore controlled by the $KQ$ tunable parameters collected in the vector $\boldsymbol{\eta}$.

The SIM is excited at the first layer by a set of $N_t$ transmitting antennas arranged in a uniform linear array (ULA) along the $y$-axis with $N_y^{tx}$ elements. Similarly, the resulting electromagnetic field is observed by $N_R$ receiving antennas (probes), also arranged in a ULA with $N_y^{prb}$ elements along the $y$ direction. Both transmitting and probing antennas are modeled as strip dipoles identical to those used within the SIM, radiating in free space.

For all considered examples, the scattering matrices $\mathbf{S}_{SS}$ used in the optimization are obtained through full-wave simulations performed with the commercial software FEKO. The adopted SIM topology represents a trade-off between computational tractability—when using Method of Moments (MoM) simulations—and physical realism.

To emulate a scenario with electromagnetically isolated SIM layers, leading to a block-diagonal structure of the matrix $\mathbf{S}$ as in \eqref{S_SS}, the ground planes supporting the dipoles in each layer are assumed to be infinite. This assumption is consistently implemented in the FEKO simulations.

Once the scattering matrix is obtained, the optimal parameter vector $\boldsymbol{\eta}$ is computed using the optimization framework described in Section~\ref{SIM_opt}. The resulting configuration defines the matrix $\mathbf{\Gamma}(\boldsymbol{\eta})$, which is then used to evaluate the SIM output matrix $\mathbf{\hat{X}}(\boldsymbol{\eta})$ according to \eqref{eq:Xhat_fast}.

Regarding the simulation setup, we consider a scenario conceptually related
to the one investigated in~\cite{An2023}, where multiple SIMs are deployed
at both the transmitter and receiver sides to enforce channel
orthogonalization and enhance the MIMO capacity. In contrast, the framework adopted in this work focuses on a simplified yet
essential configuration, where a {single} SIM is placed between a
transmitter and a receiver, as illustrated in Fig.~\ref{fig:SIMmodel}. The
role of the SIM is to manipulate the propagation environment so as to
transform the overall end-to-end channel into a nearly orthogonal one. Under
this setting, a zero-forcing–like behavior can be achieved, provided that the
resulting channel matrix has full rank.

The considered system includes $4$ transmit antennas and $4$ receive antennas,
both arranged in uniform linear arrays. The transmitter and receiver are located at distances of $10\lambda$ and $5\lambda$ from the SIM, respectively.
The SIM extends along the $y$- and $z$-axes, with the $y$-axis defining the direction of the transmit and receive linear arrays. Each layer comprises $N_y^{(q)} = 16$ elements along $y$ and $N_z^{(q)} \in [4,16]$ elements along $z$, for all $q$.

Such a geometry ensures operation in the near-field regime, where the
propagation conditions typically lead to a full-rank MIMO channel (i.e.,
rank~$4$). In this case, the effective SIM-assisted transfer function
$\mathbf{A}\,\mathbf{H}_2(\boldsymbol{\eta})\,\mathbf{\Lambda}$ is a $4\times 4$
matrix. The objective of the SIM optimization is therefore to shape this
matrix so that it becomes approximately diagonal, with dominant entries along
the main diagonal and negligible off-diagonal components. When this condition is met, the receiver can decode the transmitted data
streams independently, since the inter-stream interference is effectively
suppressed, leading to a set of parallel interference-free channels.

In the adopted configuration, each transmit antenna radiates a power of
$50$~mW. At the receiver, the disturbance is modeled as spatially white
additive Gaussian noise. 
The disturbance observed at the receiver includes both the receiver thermal noise and an equivalent contribution associated with the active devices embedded in the SIM. The receiver thermal noise variance at each receive antenna is modeled as
\begin{equation}
\sigma_{\rm rx}^2 = F k_B T_0 B,
\end{equation}
where $F$ is the receiver noise factor, $k_B$ is the Boltzmann constant, $T_0=290$~K is the reference temperature, and $B$ is the system bandwidth. Unless otherwise stated, the numerical results assume $F=6$~dB. To account for the noise generated within the active SIM, we adopt an equivalent system-level model in which the internally generated noise is assumed to scale with the overall end-to-end gain experienced by the useful signal. Accordingly, the equivalent noise variance per receive antenna is modeled as
\begin{equation}
\sigma_n^2(\boldsymbol{\eta}) = F k_B T_0 B \left(1+\frac{\|\mathbf H_{e2e}^{(S)}(\boldsymbol{\eta})\|_F^2}{M}\right),
\label{eq:noise_equiv_active}
\end{equation}
where $M$ denotes the number of receive antennas. The first term represents the conventional receiver thermal noise, whereas the second term captures the increase of the equivalent noise level associated with active SIM processing. A rigorous characterization of the internally generated noise in active SIMs, including its generation mechanisms and propagation through the multilayer structure, is beyond the scope of this work and is left for future investigations. The adopted equivalent-noise model nevertheless provides a convenient framework for assessing the sensitivity of the proposed design to active-device noise.

Different bandwidth values are considered in the numerical analysis in order
to assess the impact of noise on the achievable performance. Since the noise
power scales linearly with $B$, increasing the bandwidth results in a higher
noise level at the receiver, thus allowing us to evaluate the effectiveness
of the SIM optimization under progressively more challenging conditions.

\subsection{Results}

In this section, we report the performance of the optimized SIM configurations under different operating conditions. The results are organized into three groups, according to the parameter under investigation: bandwidth, active gain, and transfer-function diagonality.

\subsubsection{Capacity versus bandwidth}

\begin{figure}[t]
    \centering
    \includegraphics[width=1\columnwidth]{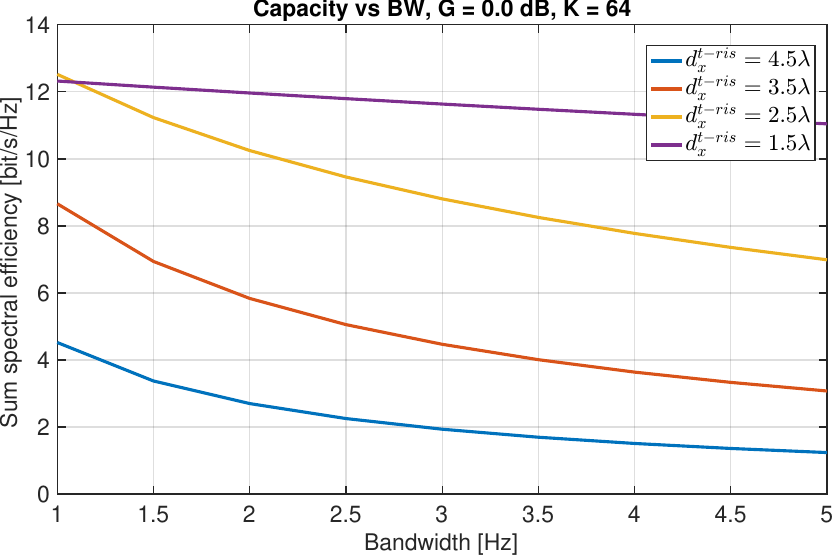}
    \caption{Sum spectral efficiency versus bandwidth for $G=0$~dB and $K=64$.}
    \label{fig:cap_bw_k64_g0}
\end{figure}

\begin{figure}[t]
    \centering
    \includegraphics[width=1\columnwidth]{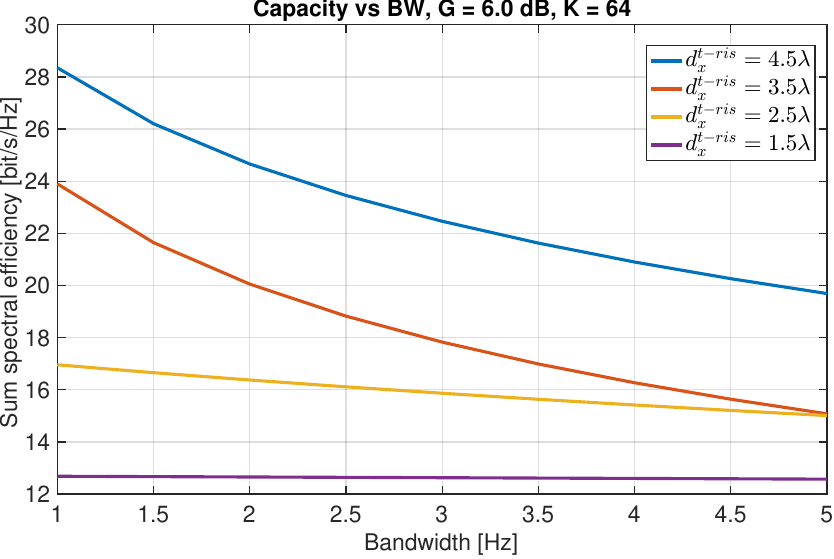}
    \caption{Sum spectral efficiency versus bandwidth for $G=6$~dB and $K=64$.}
    \label{fig:cap_bw_k64_g6}
\end{figure}

\begin{figure}[t]
    \centering
    \includegraphics[width=1\columnwidth]{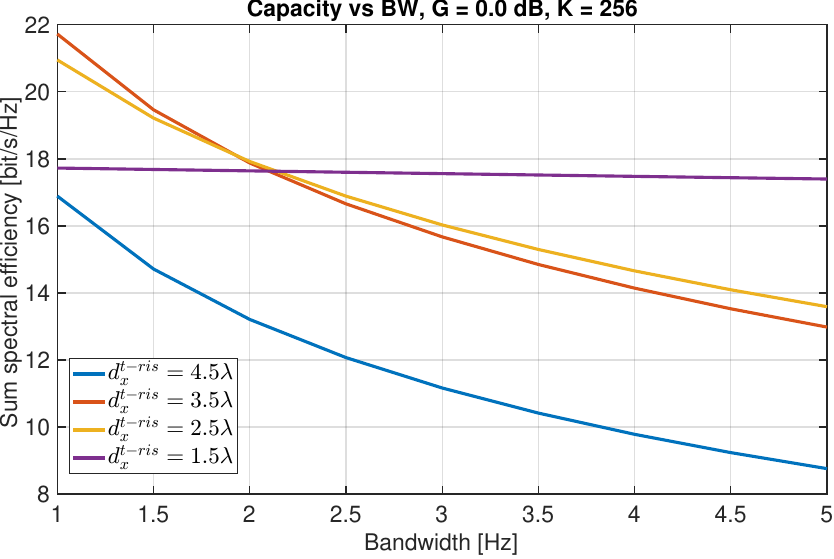}
    \caption{Sum spectral efficiency versus bandwidth for $G=0$~dB and $K=256$.}
    \label{fig:cap_bw_k256_g0}
\end{figure}

\begin{figure}[t]
    \centering
    \includegraphics[width=1\columnwidth]{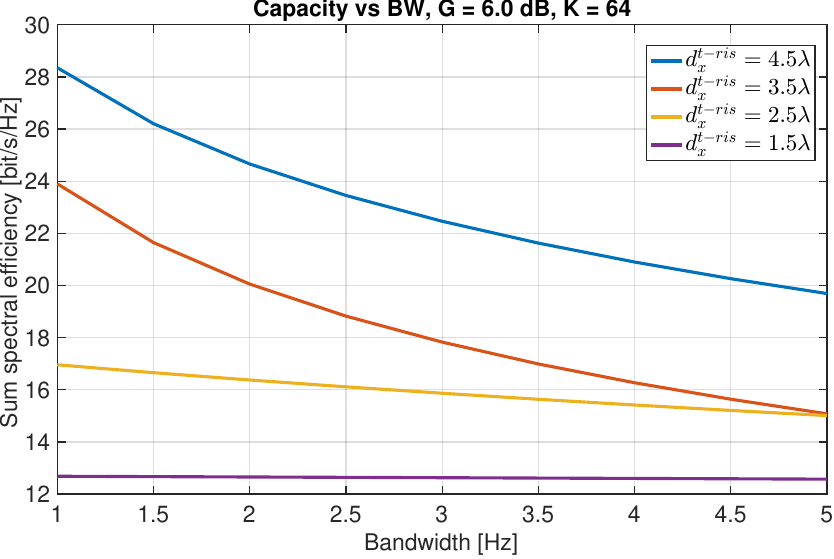}
    \caption{Sum spectral efficiency versus bandwidth for $G=6$~dB and $K=256$.}
    \label{fig:cap_bw_k256_g6}
\end{figure}

Figs.~\ref{fig:cap_bw_k64_g0}--\ref{fig:cap_bw_k256_g6} show the achieved sum spectral efficiency as a function of the bandwidth for different inter-layer spacings $d_x^{t\text{-}ris}$. Two gain configurations are considered, namely $G=0$~dB and $G=6$~dB, and two SIM sizes corresponding to $K=64$ and $K=256$ elements per layer. The increase in the SIM size is performed along the $z$-direction rather than the $y$-direction. This choice ensures a fair comparison, since the near-field condition, and therefore the rank properties of the channel, are primarily determined by the array aperture along the $y$-axis in the considered geometry. Increasing $N_y$ would instead modify the propagation regime itself, leading to a stronger near-field behavior and thus to a fundamentally different channel structure.

These figures illustrate how the system performance varies with the noise level induced by the bandwidth, as well as with the inter-layer spacing. Each curve corresponds to a different value of $d_x^{t\text{-}ris}$, highlighting the impact of propagation losses across the SIM structure.

A more detailed inspection of the first four figures reveals a clear trend. In the baseline case with $G=0$~dB and $K=64$, i.e., a relatively small SIM without active gain, reducing the inter-layer spacing is crucial. In particular, the configuration with the smallest distance ($d_x^{t\text{-}ris}=1.5\lambda$) becomes the best-performing one beyond a certain noise level. Interestingly, the performance in this case is essentially independent of the noise level, indicating that the useful signal power remains significantly larger than the noise power across all considered regimes.

On the other hand, when either the active gain $G$ is increased or the SIM size grows (i.e., $K=256$), the behavior changes. In these cases, configurations with larger inter-layer spacing tend to provide better performance. This can be attributed to the improved effectiveness of the optimization process, which is able to achieve a more accurate diagonalization of the transfer matrix, thus leading to a stronger rejection of inter-stream interference.

These observations highlight an important trade-off: active gain and SIM size can be exchanged to some extent in achieving a desired performance level. In other words, increasing the number of elements per layer can compensate for a reduced gain, and vice versa. 
It is worth emphasizing that both increasing the active gain $G$ and enlarging the SIM size $K$ also increase the equivalent noise contribution according to the model in \eqref{eq:noise_equiv_active}. Nevertheless, the numerical results consistently show a net performance improvement. This indicates that, within the considered operating range, the enhancement of the useful signal transfer provided by larger SIM apertures and higher active gains more than compensates for the associated increase in additional generated noise.

\subsubsection{Capacity versus active gain}

\begin{figure}[t]
    \centering
    \includegraphics[width=1\columnwidth]{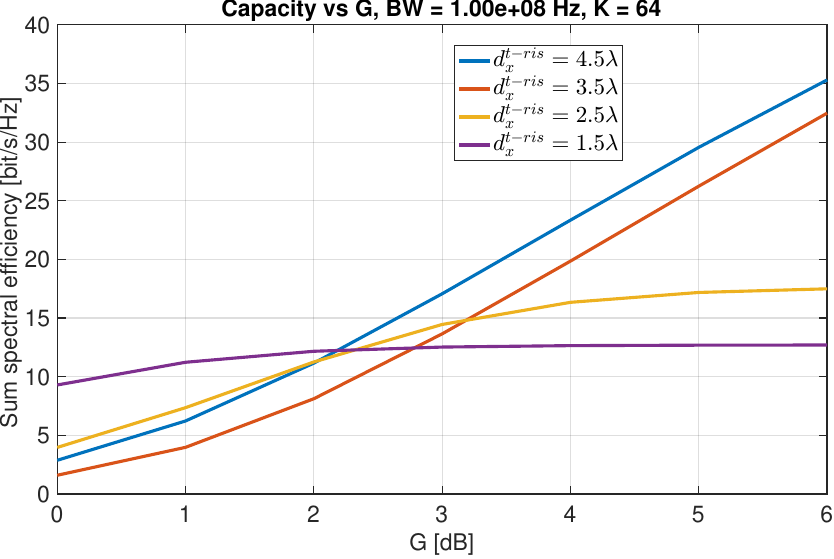}
    \caption{Sum spectral efficiency versus $G$ for minimum bandwidth and $K=64$.}
    \label{fig:cap_g_k64_bwmin}
\end{figure}

\begin{figure}[t]
    \centering
    \includegraphics[width=1\columnwidth]{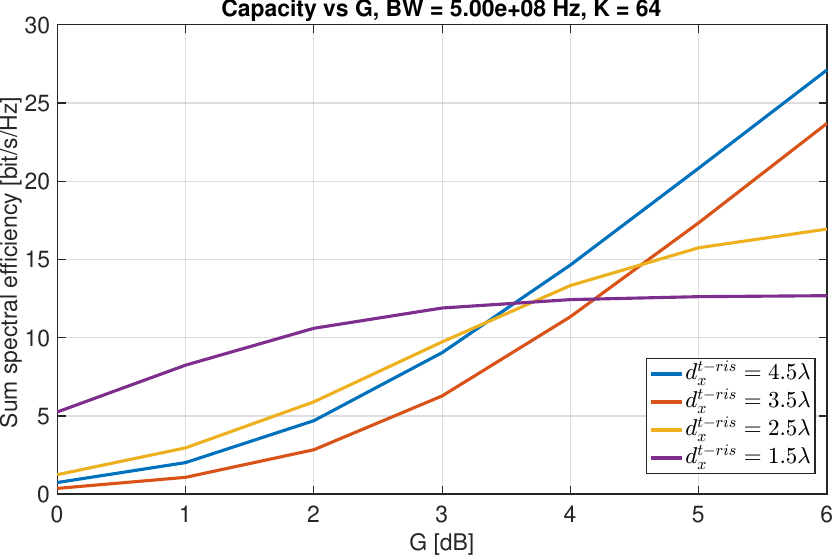}
    \caption{Sum spectral efficiency versus $G$ for maximum bandwidth and $K=64$.}
    \label{fig:cap_g_k64_bwmax}
\end{figure}

\begin{figure}[t]
    \centering
    \includegraphics[width=1\columnwidth]{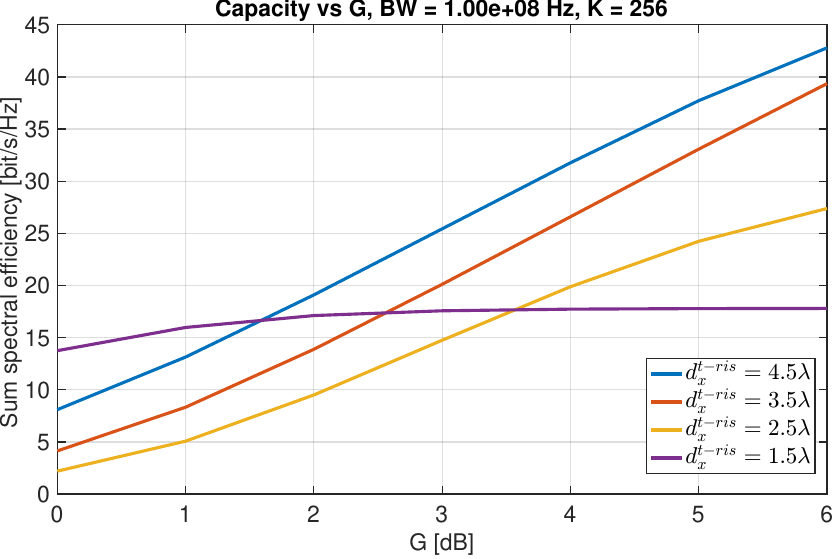}
    \caption{Sum spectral efficiency versus $G$ for minimum bandwidth and $K=256$.}
    \label{fig:cap_g_k256_bwmin}
\end{figure}

\begin{figure}[t]
    \centering
    \includegraphics[width=1\columnwidth]{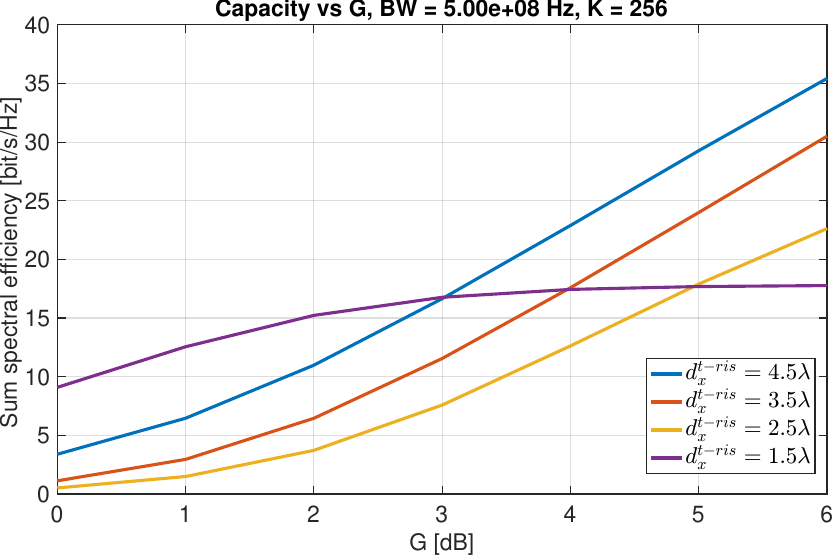}
    \caption{Sum spectral efficiency versus $G$ for maximum bandwidth and $K=256$.}
    \label{fig:cap_g_k256_bwmax}
\end{figure}

Figs.~\ref{fig:cap_g_k64_bwmin}--\ref{fig:cap_g_k256_bwmax} report the sum spectral efficiency as a function of the active gain $G$, for two representative bandwidth values (minimum and maximum considered in the simulations) and for both SIM sizes.

These plots confirm trends consistent with those observed in the previous analysis. In particular, increasing either the active gain $G$ or the SIM size $K$ leads to improved performance, as both mechanisms contribute to strengthening the useful signal relative to interference and noise. As a result, configurations with larger inter-layer spacing become progressively more favorable when $G$ and/or $K$ increase, due to the enhanced capability of the optimization process to achieve a more accurate diagonalization of the transfer matrix. The results also provide useful insight into the role of the active gain. In particular, the case $G=0$ dB corresponds to the same unilateral architecture considered throughout the paper, but without active amplification, and therefore serves as a natural reference for assessing the impact of active processing. As $G$ increases, the active interconnection networks progressively compensate for the attenuation associated with inter-layer propagation, resulting in a stronger end-to-end transfer and improved spectral efficiency.

\subsubsection{Transfer-function diagonality}

\begin{figure}[t]
    \centering
    \includegraphics[width=1\columnwidth]{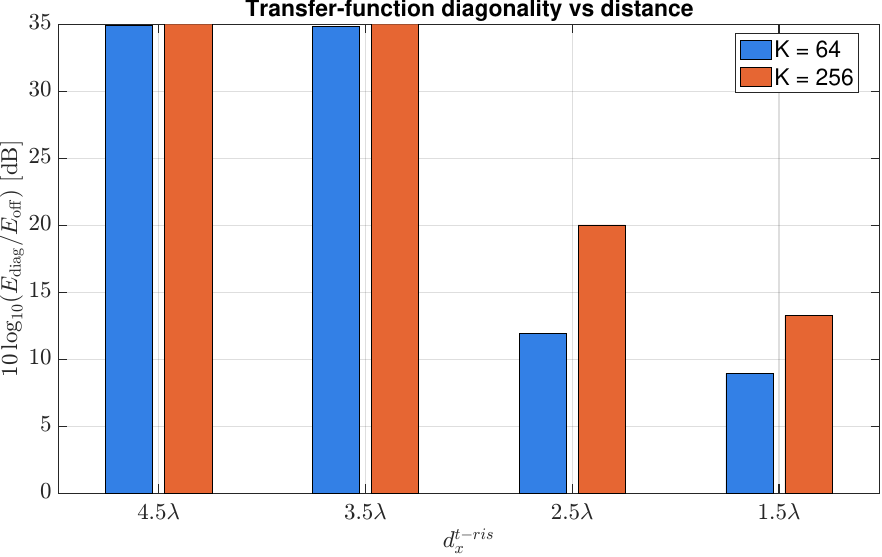}
    \caption{Transfer-function diagonality versus inter-layer spacing for $K=64$ and $K=256$.}
    \label{fig:diag_bar}
\end{figure}

Finally, Fig.~\ref{fig:diag_bar} reports the diagonality metric of the transfer function as a function of the inter-layer spacing, for both SIM sizes.

The metric represents the ratio between the energy concentrated on the diagonal of the transfer matrix and the energy on the off-diagonal elements, thus providing a measure of how accurately the SIM realizes the desired input-output mapping. It is worth emphasizing that this metric is evaluated in the absence of noise, and therefore reflects only the effectiveness of the optimization process.

As shown in the figure, larger inter-layer spacing enables a more accurate optimization of the objective function, allowing the algorithm to reach the target threshold of $35$~dB, which corresponds to the stopping criterion of the iterative procedure. In contrast, for smaller spacings (e.g., $d_x^{t\text{-}ris}=2.5\lambda$ and $1.5\lambda$), this threshold is not achieved, although the resulting diagonality remains relatively high. Note that Figure \ref{fig:diag_bar}  provides additional insight into the trends observed in previous results by directly quantifying the quality of the achieved diagonalization. In particular, the reduction of the transfer-function error for larger inter-layer spacings confirms that the corresponding performance gains originate from a more accurate approximation of the desired diagonal end-to-end transfer matrix.

These results highlight a trade-off already discussed in the previous sections: configurations with smaller inter-layer spacing generally provide higher output signal power, but may limit the capability of the optimizer to fully diagonalize the transfer function. Conversely, larger spacing improves the achievable diagonality, at the expense of increased propagation losses across the SIM.

Finally, Table~\ref{tab:convergence} summarizes the convergence behavior of the proposed gradient-based optimization algorithm for the considered inter-layer spacings and for $K = 64$. All simulations were performed using a maximum of 10\,000 iterations.
\begin{table}[t]
\caption{Representative convergence behavior of the proposed optimization algorithm for $K = 64$.}
\label{tab:convergence}
\centering
\begin{tabular}{c|c}
\hline
Inter-layer spacing & Iterations to convergence \\
\hline
$4.5\lambda$ & $\approx 1000$ \\
$3.5\lambda$ & $\approx 2000$ \\
$2.5\lambda$ & $>10\,000$ \\
$1.5\lambda$ & $>10\,000$ \\
\hline
\end{tabular}
\end{table}
As can be observed from Table~\ref{tab:convergence}, the convergence behavior strongly depends on the inter-layer spacing. For $4.5\lambda$ and $3.5\lambda$, the proposed algorithm typically reaches the stopping criterion after 1000-2000 iterations, whereas for $2.5\lambda$ and $1.5\lambda$ convergence is not achieved within the maximum number of allowed iterations. These results indicate that strongly coupled SIM configurations give rise to a more challenging optimization landscape and may benefit from more sophisticated optimization strategies, whose investigation is left for future work.
\section{Conclusion}
We proposed a multi-port S-parameter framework for SIMs with unilateral active interconnections, which combines electromagnetic accuracy with a feed-forward structure enabling a recursive cascade representation of the system. This formulation allows for a significant reduction in computational complexity and leads to an efficient gradient-based optimization algorithm. Numerical results highlighted the trade-offs among inter-layer spacing, active gain, and SIM size, showing that signal attenuation and optimization capability can be effectively balanced. The proposed approach provides a bridge between realistic electromagnetic modeling and simplified cascade representations, offering a practical tool for the design of SIM-based communication systems.


\appendices

\section{Inter-Layer Isolation Assessment}
\label{app:isolation}

In this appendix, we provide a quantitative assessment of how well
a practical, finite-size implementation approximates the ideal
inter-layer isolation obtained with the infinite ground plane model
used in the full-wave simulations.

\subsection*{Simulation Setup}
\begin{figure}[t]
	\centering
	\includegraphics[width=1\columnwidth]{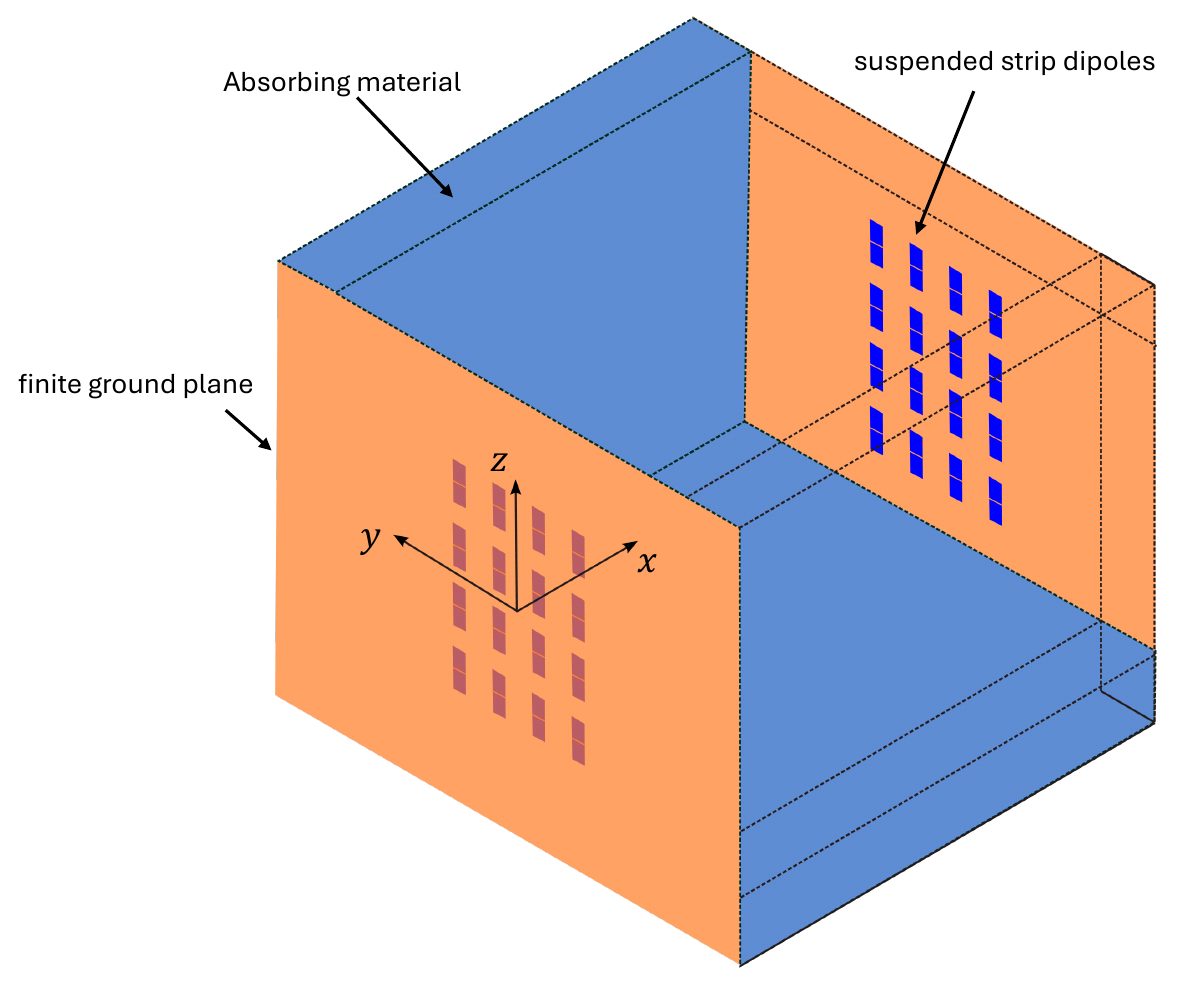}
	\caption{Open view of the CAD model used in full-wave simulations.
	The two finite ground planes are visible, each supporting a
	$4\times 4$ strip-dipole array. Two of the four lateral absorbing
	panels (Laird AN73(front), blue solids) are also shown.}
	\label{fig:boxgeom}
\end{figure}
The practical geometry, implemented and simulated in FEKO using the
Method of Moments (MoM), consists of two finite ground planes facing
each other at a distance $d_x^{t\text{-}ris}$, each supporting a
$4 \times 4$ uniform planar array of suspended strip dipoles.
The array size is kept small to limit the computational effort.
The two arrays represent the transmit array of layer $q$ (Array~A,
ports 1--16) and the receive array of layer $q+1$ (Array~B,
ports 17--32), respectively.
To suppress cavity modes and emulate the behavior of infinite ground
planes, the four lateral walls of the enclosure are lined with a
commercial flat absorbing material (Laird AN73front, nominal thickness
$t_f = 10\ \mathrm{mm}$), backed by metallic sheets connected to
the finite ground planes.
Fig.~\ref{fig:boxgeom} shows an open view of the geometry, where
the two finite ground planes and two of the four absorbing panels
are visible.
This configuration is similar to the practical inter-layer isolation
approach described in~\cite{Liu2022}.
Simulations are carried out for four values of the ground-plane
separation: $d_x^{t\text{-}ris} \in \{1.5,\, 2.5,\, 3.5,\, 4.5\}\,\lambda_0$,
matching the spacing values used in the main results.
For reference, the absorbing enclosure simulation at
$d_x^{t\text{-}ris} = 4.5\lambda_0$ requires approximately 5~hours
of computation and 176~GB of RAM (Intel Core Ultra 9 285K, 24-core, 3.7~GHz, 256~GB system RAM), compared to about 66~seconds and
50~MB of RAM for the infinite ground plane case, which exploits the
FEKO closed-form Green's function for infinite conducting planes.

\subsection*{Metrics and Results}
Two complementary metrics are used to compare the S-parameter matrices
obtained with infinite ground planes and with the absorbing
enclosure (Box+AN73 absorber).
The first metric is the root-mean-square (RMS) amplitude of the
mutual S-parameters between Array~A and Array~B, defined as
\begin{equation}
\text{RMS}
=
\sqrt{\frac{1}{K^2}\sum_{i,j}|S_{ij}|^2},
\end{equation}
where the sum is taken over all cross-coupling entries between the
two arrays. This metric captures the overall mutual coupling energy
between the two arrays.
The second metric is the Pattern Distance (PD), which measures the
element-by-element mismatch between the two normalized S-matrices:
\begin{equation}
\text{PD}
=
\left\|
\frac{\mathbf{S}_{\text{Box}}}{\|\mathbf{S}_{\text{Box}}\|_F}
-
\frac{\mathbf{S}_{\text{NoBox}}}{\|\mathbf{S}_{\text{NoBox}}\|_F}
\right\|_F.
\end{equation}
By normalizing each matrix to its Frobenius norm before computing
the difference, PD isolates the spatial redistribution of the
coupling pattern from any difference in the overall amplitude level.

\begin{figure}[t]
    \centering
    \includegraphics[width=1\columnwidth]{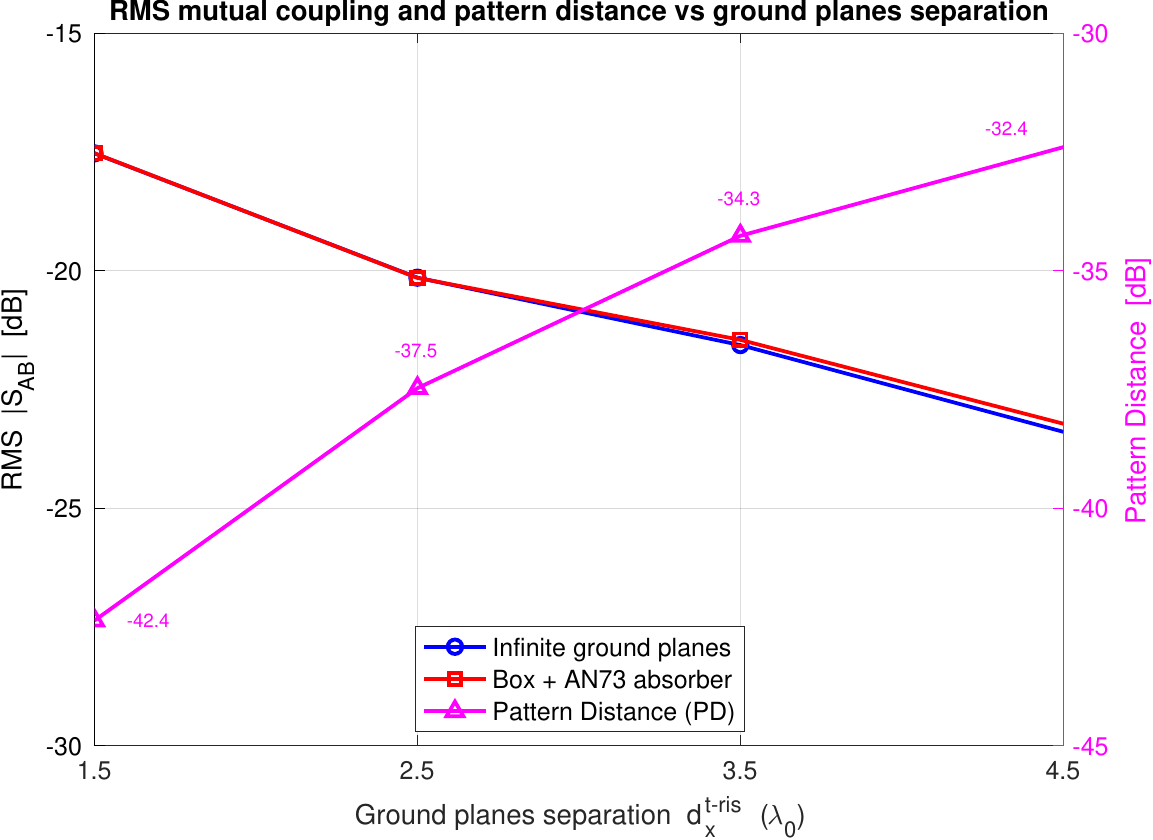}
    \caption{RMS amplitude of the mutual S-parameters (left axis,
    solid curves) and Pattern Distance PD (right axis, dashed curve)
    between Array~A and Array~B as a function of the ground planes
    separation $d_x^{t\text{-}ris}$ $(\lambda_0)$, for the infinite
    ground plane (NoBox) and the absorbing enclosure (Box) configurations.}
    \label{fig:rms_pd_combined}
\end{figure}

Fig.~\ref{fig:rms_pd_combined} shows both metrics as a function of
$d_x^{t\text{-}ris}$.
The RMS curves (left axis) obtained with the two configurations are
nearly superimposed across all tested spacings and decrease
monotonically as $d_x^{t\text{-}ris}$ increases, confirming that
the two geometries carry essentially the same mutual coupling energy
at every separation distance.
The PD values (right axis) remain in the range $-42$ to $-32$~dB
across all spacings, confirming that the spatial distribution of
the coupling pattern is only marginally affected by the practical
enclosure.
The moderate increase of PD with $d_x^{t\text{-}ris}$ is consistent
with the onset of residual cavity modes due to imperfect absorption
by the flat absorber walls at grazing incidence, which redistribute
coupling among array elements without significantly altering the
total coupled power.

Together, the two metrics confirm that the metallic enclosure with
Laird AN73front absorber is a reliable practical approximation of the
infinite ground plane model, with a Pattern Distance consistently
below $-32$~dB across all the inter-layer spacings considered in
this work.

\section*{Acknowledgments}
			
\bibliographystyle{IEEEtran}
\bibliography{biblio_ActiveSim}
\end{document}